\documentclass{chep01}
\RequirePackage{url}
\begin{document}
\begin{center}
\Large\bf{XML for Detector Description at GLAST\footnote{Work supported 
by the Department of Energy, contract DE-AC03-76SF00515}}
\vskip0.3cm
\end{center}  
J. Bogart\ \ (SLAC, Menlo Park)\\
D. Favretto, R. Giannitrapani\ \ (Universit\`{a} di Udine, Udine)


\begin{abstract}
The problem of representing a detector in a form which is accessible to a
variety of applications, allows retrieval of information in ways which are
natural to those applications, and is maintainable has been vexing physicists
for some time.  Although invented to address an entirely different problem
domain, the document markup meta-language XML is well-suited to
detector description.
This paper describes its use for a GLAST detector.\cite{LATref}

\end{abstract}


\section{Introduction}
For any experiment an accurate description of its apparatus
is essential.  Some features of HEP experiments make this goal
especially difficult to achieve:
\begin{itemize}
\item Detectors are large and complex.
\item Experiments tend to be long-lived.  Meanwhile prototypes, actual
and simulated, must be supported; production detectors undergo 
upgrades; software technology evolves.
\item Several applications need access to the detector description.
\end{itemize}
Hence the scheme used for detector description must be flexible
and versatile.  Any information potentially of interest to more
than one application, in particular \emph{all geometric parameters and
relationships, must exist as data, not code}.

\subsection{XML}
XML\cite{XMLspec} was originally intended
as a means of designing web document types so that any part of the
content of the document,
not just formatting information, could
be tagged and thus made readily available to programs.
However, it quickly became
clear that such ``documents'' could include databases or structured
messages as well as more typical document content. 
As a result of its wide appeal, including many
commercial applications, high-quality tools for XML
soon became available and continue to evolve rapidly.  
XML is not only a suitable
vehicle for detector description now, but is well-supported and likely
to remain so for years to come.  It is therefore not surprising that
XML has gained broad popularity in the HEP community;
many experiments have adopted it as their geometry
persistency mechanism\footnote{In the last two years three specialized
workshops have been dedicated to discuss this topic among different
experiments; see, for example \cite{xmlwork}.}

\subsection{GLAST LAT Specifics}
Our task is to provide a description for the
Large Area Telescope (LAT), the primary instrument of the two composing
the GLAST observatory scheduled for launch in 2006.
It is a small, relatively simple HEP detector designed to detect gamma
rays over a large energy range.
Active volumes are practically all boxes,
channel counts in two of three subdetectors are modest,
and tolerances are generally less
stringent than for large HEP experiments.

In other respects the LAT detector and the environment
in which its software is being developed are typical of HEP.
The usual collection of programs---simulation, reconstruction, 
calibration procedures, etc.---are 
all necessary.  Most are
C++ applications being structured using the Gaudi framework which
must run on multiple platforms (Windows, Linux, Solaris).
The architecture must support laptops
disconnected from the net as well as less constrained computing environments.
Tools have to survive for the lifetime of the project, at least a decade.

\section{Implementation}
There are three major pieces to our XML description schema, 
currently implemented as a DTD known as GDD,
and to the software interpreting it.
At its heart is a method for describing geometric solids
and their relative positions.  
The second piece, a mechanism for definition, manipulation and
substitution of symbolic constants, was designed to
aid in maintenance and comprehensibility of the geometry description.
Finally, to divorce all potentially shared parts of detector description from
any single application, GDD includes facilities to
define and manipulate identifiers associated with
parts of the detector. 

\subsection{XML Parser}
We use facilities of XML
to enforce constraints on the detector description whenever possible,
hence we need a validating parser.  It must be compatible with C++
since vital applications such as simulation are written in C++.
Furthermore, parser support for the Document Object Model (DOM)\cite{DOM}
interface 
saves us from having to design our own generic XML in-memory 
representation\footnote{A larger experiment might balk at the memory
usage of the DOM, but it is unlikely to be an issue for us.}.
We have chosen
the open-source Xerces-c parser\cite{xerces}.  It has all required features,
and additionally has a large, supportive community of users and
developers.  

\subsection{Geometry}
\subsubsection{XML description}
The plethora of possible clients of the GLAST geometry has lead to a
quite general design for those XML elements in GDD dedicated to 
geometry, mainly inspired by AGDD\cite{AGDDref}, 
the ATLAS DTD, which in turn owes much to simulators such as GEANT.

The DTD defines \emph{volumes} and \emph{positioning elements}.
Volumes may be geometric primitives, like boxes or tubes, or
composites constructed of several other (primitive or composite)
volumes, nested to arbitrary depth.
Positioning elements describe the relative positions
of child volumes within their parent.  

\subsubsection{detModel and clients}
In order to isolate client applications from XML itself, we have
developed a C++ layer called \emph{detModel}, a 
class hierarchy used by GLAST applications to access the
geometry description of the detector stored in an XML file. Any
software component needing 
geometrical information accesses it via detModel; the details of
the XML description and the parser interface are hidden.

To develop this package, our analysis started from GDD and
its hierarchical structure.  XML is not object oriented, but there is
a natural mapping from GDD elements to classes connected by
 ``has-a'', ``is-a'' and ``use-a'' relations, suggesting a
possible hierarchy for these classes. This internal structure
is quite independent of client nature and access,
for which
we have added a series of management
classes and query functionalities on the geometric hierarchy. From the
start we have
tried to implement a certain degree of modularity; in
this way every client can instantiate and access only the part of GDD
that it really needs.

Throughout the design and implementation of detModel we used 
design patterns \cite{pattern} extensively. 
In particular the manager class \emph{GDDmanager},
the gateway to the full hierarchy, is a \emph{singleton};
moreover we have separated the construction of the geometry representation
from the representation itself by adopting a \emph{builder} pattern. In
this way, for example, we have decoupled detModel (and all its
clients) from the specific XML parser used, 
leaving open the option to change it in future.
We also adopted the \emph{visitor} pattern as the basic mechanism
for accessing the full hierarchy; every client needs only to implement
a concrete visitor, implementing whatever functionalities it
requires. For clients that do not need to access the full hierarchy,
but only a part of it we
have also implemented some basic direct query mechanisms.

With both the visitors mechanisms and the direct query methods of
the \emph{GDDmanager} public interface it is quite easy to 
quickly implement
specific requirements for a broad collection of clients. 
We have
already developed a VRML client for graphical
representation of the geometry, a GEANT4 client that can use detModel
to instantiate the geometry in the Monte Carlo simulation (including
sensitive detector information contained in GDD) and a prototype ROOT
visitor to access the geometric information of GLAST inside analysis
and event display programs. See \cite{gallery} for sample images produced
by these clients. For the future we are working on a HepRep
client of detModel for some possible links with WIRED and JAS clients
and a GISMO client (the Monte Carlo toolkit currently used by the GLAST
collaboration). 

\subsection{Constants}
In describing even a relatively small detector like the LAT naively
with XML, one soon mourns the absence of
standard conveniences, such as support for constant names
and arithmetic.  Dimensions are often repeated; offsets are commonly
calculated from other values.  If all such values are entered as 
literal constants the description quickly becomes unmaintainable.

\subsubsection{XML definitions}
Although XML has no built-in arithmetic one can define elements
to \emph{specify} the arithemetic
to be done by co-operating applications.  We define XML
elements for ``primary'' and ``derived'' constants, for references
to constants, and for standard
arithmetic operations.  Primary constants have only a name, a 
value, type information, and a descriptive comment.  
Derived constants may have element content
consisting of arithmetic elements and references to other
constants. Other XML elements, such as those describing volumes
of a particular shape, have been modified to accept either a literal
value for, e.g., a radius, or a reference to a constant.
Constant names are IDs in the XML sense and references to
them are IDREFs, so validating parsers enforce uniqueness
of such names and verify that all references resolve properly.

\subsubsection{Supporting software}
Utility C++ classes have been written to evalute derived constants,
optionally replacing their element content with a value attribute,
and to substitute a value for a reference to a constant, doing the
evaluation first if necessary.  These utilities may be invoked 
directly by programs such as Simulation.  A stand-alone program
using these utilities takes an arbitrary XML document as input
and outputs an equivalent document with all expressions evaluated
and all references replaced with values.  For production running
this preprocessing saves resources, most notably the memory required
for the DOM representation, which is considerably smaller and
less nested for the preprocessed document.

The constants have interest in their own right, both to programs
which need access to a particular parameter apart from the complete
geometry description, and to humans in search of a reliable source
of documentation.  C++ classes have been written to ``serve'' the
constants via both visitor and query interfaces, as has a stand-alone 
visitor client which produces html tables of constants, readily
accessible to the full collaboration for review.

\subsection{Identifiers}
Several clients of detector description need a way to identify
individual volumes, both for internal use and to label
information, such as hits, which may be generated by one 
application and read by another.  
No single labeling scheme can gracefully accomodate all clients.
An analysis client may only be concerned with read-out volumes 
whereas Simulation must be aware of all volumes.
GDD supports multiple labelings and, where feasible,
transforms between them.

\subsubsection{XML definitions}
Positioning elements may include an \emph{idField}, which is
composed of a fieldname and value.
An \emph{identifier} for a volume is the concatenation of the values of
its own and ancestor idFields.  

In order to validate an identifier or manipulate
identifiers as a group, we use an
\emph{identifier dictionary}.\footnote{The concept and name are due
to members of the ATLAS AGDD Working Group.}  A dictionary
specifies legal sequences of fieldnames and constraints on values
in a particular context.  Dictionaries may be used to constrain
sets of identifiers (such as readout identifiers)
other than those  coming from positioning elements.
Conversions taking identifiers from one dictionary 
to another are specified using an
\emph{id converter}.  
Values occurring in dictionaries or converters 
can be literals or references to be resolved by substitution.

\subsubsection{Supporting software}
GDD imposes a hierarchical structure 
on the set of identifiers belonging to a given dictionary, but does
not enforce certain uniqueness properties.  Id dictionary classes
have a validation function to ensure, e.g., that any allowed
sequence of field values has a uniquely-associated sequence of fieldnames.
Individual identifiers may be checked for consistency with a dictionary.

\section{Status and Conclusions}
The current description of the LAT using GDD,
including a compatible id dictionary, describes all active and some 
structural components.  Remaining components will be 
added as needed.   Except for id conversion, all
fundamental services needed to interpret a GDD document have been
implemented.  Several key clients are up and running; others are in the
works.

As the LAT description grows in size and complexity, we need to monitor
potential resource bottlenecks, particularly memory.  
Maintenance of the description files is more immediately troublesome.  
The ID/IDREF mechanism is heavily used in GDD, which leads to
problems in name selection and management: ID values must be a
unique within the document.  A move to XML Schema,
a newer alternative to DTDs, would help.\footnote{XML Schema has 
other advantages which space does not permit us to describe.}
GDD documents are fragile in the sense that
small change in the design of the detector can entail large changes
in the document.  However, if the information in the document were
kept in code  instead such changes would be significantly more disruptive.
In summary, using XML to describe detectors, while not ideal,
appears to be workable, and probably significantly superior to 
older technologies.

\end{document}